\documentclass[a4paper]{article}

\usepackage{amssymb, amsmath, graphicx}

\begin{document}

\title{A comparison of a few numerical schemes for the integration of stochastic differential equations in the Stratonovich interpretation}

\author{
David Garc\'\i a-\'Alvarez$^{\ \mathrm\ast}$\\
\\
\textit{Georges Lema\^\i tre Centre for Earth and Climate Research} \\
\textit{Universit\'e Catholique de Louvain}\\
\textit{Chemin du Cyclotron, 2 - BE-1348 Louvain-la-Neuve - Belgium}\\
$^{\mathrm\ast\ }$\texttt{david.dga@gmail.com}
}

\date{21 February 2011}

\maketitle

\begin{abstract}
Three schemes, whose expressions are not too complex, are selected for the numerical integration of a system of stochastic differential equations in the Stratonovich interpretation: the integration methods of Heun, Milstein, and derivative-free Milstein. The strong (path-wise) convergence  is studied for each method by comparing the final points after integrating with $2^n$ and $2^{n-1}$ time steps. We also compare the time that the computer takes to carry out the integration with each scheme. Putting both things together, we conclude that, at least for our system, the Heun method is by far the best performing one.
\end{abstract}

\section{Introduction}

\subsection{Stochastic differential equations}

Let us suppose that we are dealing with a system with deterministic dynamics, on which external noise is acting. That means that, in case we could remove the noise (by switching it off if possible, by isolating the system, etc), its equations of motion would be purely deterministic:

\begin{equation}\label{eq:deterministic}
\mathrm{d} \boldsymbol{X}(t) = \boldsymbol{f} ( \boldsymbol{X}(t) , t) \, \mathrm{d}t.
\end{equation}

Now, let us consider that Gaussian white noise is acting on the system. To include its effects, we add to (\ref{eq:deterministic}) a term $\boldsymbol{g} ( \boldsymbol{X}(t) , t) \, \mathrm{d}\boldsymbol{W}(t)$:

\begin{equation}\label{eq:stochastic}
\mathrm{d} \boldsymbol{X}(t) = \boldsymbol{f} ( \boldsymbol{X}(t) , t) \, \mathrm{d}t + \boldsymbol{g} ( \boldsymbol{X}(t) , t) \, \mathrm{d}\boldsymbol{W}(t),
\end{equation}
where the functions of $\boldsymbol{g}$ are determined by some physical considerations, and the components of $\boldsymbol{W}$ are independent Wiener processes. A Wiener process  \cite{kloeden_platen} 
 is a Gaussian stochastic process, almost surely continuous, where non-overlapping increments are independent, and with

\begin{eqnarray}\label{wiener_process}
\mathrm{(i)} & & W(0) = 0,\nonumber\\
\mathrm{(ii)} & & \left< W(t) \right> = 0,\nonumber\\
\mathrm{(iii)} & & \mathrm{Var} (W(t) - W(s)) = t-s.
\end{eqnarray}

\subsection{Ito and Stratonovich interpretations}

Equation (\ref{eq:stochastic}) can be written as

\begin{equation}\label{eq:stoc_integrated}
 \boldsymbol{X}(t) =  \boldsymbol{X}(t_0)+\int_{t_0}^t \boldsymbol{f} ( \boldsymbol{X}(s), s) \, \mathrm{d}s +\int_{t_0}^t \boldsymbol{g} ( \boldsymbol{X}(s) , s) \, \mathrm{d}\boldsymbol{W}(s),
 \end{equation}
 %
%
An issue arises when defining the integral over the Wiener processes. If we take a partition $\{ t_0,\dots, t_N\}$ of the interval $[t_0, t]$ (where $t_N = t$), and being $\tau_i$ a value between $t_{i-1}$ and $t_i$, $\tau_i = (1-\lambda)t_{i-1} + \lambda t_i$, we would define the integral of any function $\Phi$ as \cite{burrage}

\begin{equation}
\int_{t_0}^t \Phi(\boldsymbol{X}(s), s)\, \mathrm{d}W(s) = \lim_{\Delta\to 0} \sum_{i=1}^N\ \Phi(\boldsymbol{X}(\tau_i), \tau_i)[W(t_i)-W(t_{i-1})],
\end{equation}
where $\Delta = \mathrm{max}\{t_i-t_{i-1} | i\}$. The issue is that the limit above is different for different choices of $\lambda$. The most common interpretations are the Ito scheme, in which the function is evaluated at the left-hand endpoint of each sub-interval ($\lambda=0$), and the Stratonovich scheme, in which the function is evaluated at the midpoint of each sub-interval ($\lambda = 1/2$). We usually write the Stratonovich calculus with a circle before $\mathrm{d}W$.

The solutions to the stochastic differential equations are different for both interpretations, given that the stochastic integrals have a different value for each one. For example, the integral bellow in the Ito calculus is \cite{kloeden_platen}, 

\begin{equation}\label{integral_ito}
\int_{t_0}^t  W(s)\, \mathrm{d}W(s) = \frac{1}{2} [W (t)^2 - W (t_0)^2] - \frac{1}{2}(t-t_0),
\end{equation}
while for Stratonovich calculus,

\begin{equation}\label{integral_stratonovich}
\int_{t_0}^t  W(s)\circ \mathrm{d}W(s) = \frac{1}{2} [W (t)^2 - W (t_0)^2].
\end{equation}

Chosen one interpretation, one can always do the calculus in the other interpretation by modifying $\boldsymbol{f}$ with the \textit{Ito-Stratonovich drift correction formula} 
\cite{kloeden_platen,
risken
}: the two equations bellow

\begin{eqnarray}
\mathrm{d} \boldsymbol{X}(t) & =& \boldsymbol{f} ( \boldsymbol{X}(t) , t) \, \mathrm{d}t + \boldsymbol{g} ( \boldsymbol{X}(t) , t) \, \mathrm{d}\boldsymbol{W}(t)\\
\mathrm{d} \boldsymbol{X}(t) &=& \underline{\boldsymbol{f}} ( \boldsymbol{X}(t) , t) \, \mathrm{d}t + \boldsymbol{g} ( \boldsymbol{X}(t) , t) \circ \mathrm{d}\boldsymbol{W}(t)
\end{eqnarray}
have the same solution if their drifts fulfil the relationship

\begin{equation}\label{drift_correction}
\underline{f}_{\ i} ( \boldsymbol{X}(t) , t) = f_i ( \boldsymbol{X}(t) , t) - \frac{1}{2} g_{kj} ( \boldsymbol{X}(t) , t)\, \frac{\partial}{\partial x_k} g_{ij} ( \boldsymbol{X}(t) , t).
\end{equation}


When the equations (\ref{eq:stochastic}) model a physical system, the ``white noise" $\mathrm{d}W$ is actually an idealization of noise with a small correlation time: the autocorrelation function of $\mathrm{d}W(t)$ is not really a delta function, but a sharply peaked one; i.e., its correlation time is small, but positive. Therefore, the stochastic function $\mathrm{d}W(t)$ is actually not singular,  and the stochastic differential equations (\ref{eq:stochastic}) have a well-defined solution. In the limit of correlation time of $\mathrm{d}W(t)$ going to zero, such solution tends to the integration of the equations with the Stratonovich interpretation (see more details and a discussion in chapter IX.5 of \cite{vankampen}, \textit{Discussion of the Ito-Stratonovich dilemma}). That is why we will use the Stratonovich interpretation.

\section{Integration methods}\label{sec:integration_methods}

We will consider three light-weight (in terms of the complication of the formulae) numerical integration methods leading to the Stratonovich interpretation. They all require using a sample of the discretized Wiener process at each integration step. Let $h$ be the length of the time step used for the numerical integration. Therefore, according to the properties of the Wiener process (\ref{wiener_process}),  we need a sample of independent random variables $\xi_i(t_n)$ at each of the values of the time $\{t_n\}$ in the discretization, distributed as $\sqrt{h}\, \mathcal{N}(0,1)= \mathcal{N}(0,h)$, where  $\mathcal{N}(a,\sigma^2)$ stands for the normal distribution with mean $a$ and variance $\sigma^2$.

\subsection{Heun method}

One of the simplest discretization schemes leading to the Stratonovich interpretation is the Heun method 
\cite{
nowak
, ruemelin,
burrage,
noise_spatially
}.
This is a predictor-corrector method: given the value of $\boldsymbol{X}$ at a time $t_n$ of the discretization, we first obtain the predictors, or supporting values, with the Euler integration scheme

\begin{equation}\label{heun_support}
\bar{x}_i (t_{n+1}) = x_i (t_n) + f_i (x(t_n), t_n)\, h + g_{ij} (x(t_n), t_n)\, \xi_j(t_n),
\end{equation}
where $t_{n+1} = t_n+h$. Then, we obtain $\boldsymbol{X}(t_{n+1})$ as

\begin{eqnarray}\label{heun}
x_i (t_{n+1}) &=& x_i (t_n) + \frac{1}{2}\left[ f_i (x(t_n), t_n)+ f_i (\bar{x} (t_{n+1}), t_{n+1})\right]\, h\nonumber\\
&&\quad +  \frac{1}{2}\left[ g_{ij} (x(t_n), t_n)+ g_{ij} (\bar{x} (t_{n+1}), t_{n+1})\right]\, \xi_j(t_n).
\end{eqnarray}
Note that we are using Einstein's tensor convention of summing over repeated indices.

\subsection{Milstein scheme}

The Milstein scheme requires the use of the derivatives of the diffusion coefficients. In the general case, it reads \cite{kloeden_platen
}

\begin{equation}\label{milstein_general}
x_i (t_{n+1}) = x_i (t_n) + f_i\, h + g_{ij}\, \xi_j + g_{lj}\frac{\partial g_{ik}}{\partial x_l}\, J_{(j, k)},
\end{equation}
where all the terms at the right hand side of (\ref{milstein_general}) are taken at time $t_n$, and $J_{(j, k)}$ is a multiple Stratonovich integral

\begin{equation}\label{multiple_strat_integral}
J_{(j, k)}= \int_{t_n}^{t_{n+1}}\left( \int_{t_n}^{s_1} \mathrm{d}W_j (s_2) \right) \mathrm{d}W_k (s_1)
\end{equation}
(when using the Ito interpretation, the corresponding multiple Ito integral is defined in the same way as in (\ref{multiple_strat_integral}), but with the Ito interpretation, and written as $I_{(j, k)}$.) 

These integrals cannot be easily expressed in terms of the increments $\xi_j$ and  $\xi_k$ of the components of the Wiener process. Nevertheless, we can consider the particular case of diagonal noise, where each component of the system has its own, independent, noise, i.e.: there are as many independent Wiener processes (number of components of $\boldsymbol{W}$) as the number of variables in the system (the number of components of $\boldsymbol{X}$), each component $x_i$ of $\boldsymbol{X}$ is affected only by the corresponding component $W_i$ of the Wiener process, and the diagonal diffusion coefficient $g_{ii}$ depends only on $x_i$ and maybe on time (but not on the other components of $\boldsymbol{X}$). In such a case, $g_{lj}$ in (\ref{milstein_general}) is not equal to zero only where $l=j$, $g_{ik}$ only where $i=k$, and ${\partial g_{ii}}/{\partial x_l}$ only where $i=l$. Putting all things together, the last term in (\ref{milstein_general}) is equal to

\begin{displaymath}
g_{ii}\frac{\partial g_{ii}}{\partial x_i}\, J_{(i, i)};
\end{displaymath}
(it is clear that we are not summing over $i$ in the expression above , given that $i$ is an external index in (\ref{milstein_general}); we will omit this remark wherever it would be easy to figure out over which repeated indices we are not summing up.)
Regarding the last factor in the product above,  (\ref{multiple_strat_integral}) becomes for $j=k=i$, 

\begin{eqnarray}
J_{(i, i)}&=& \int_{t_n}^{t_{n+1}} [W_i (s_1)-W_i(t_n)]\circ \mathrm{d}W_i (s_1)\nonumber\\
& =& \int_{t_n}^{t_{n+1}} W_i (s_1)\circ \mathrm{d}W_i (s_1) - W_i(t_n)\int_{t_n}^{t_{n+1}}\circ\, \mathrm{d}W_i (s_1)\nonumber\\
&=& \frac{1}{2}[W_i(t_{n+1})^2 - W_i(t_n)^2] - W_i(t_n) [W_i(t_{n+1})-W_i(t_n) ] \nonumber\\
&=&  \frac{1}{2}[W_i(t_{n+1}) - W_i(t_n)]^2.
\end{eqnarray}
where (\ref{integral_stratonovich}) has been used at the third equality. Therefore, for the numerical integration, we set \cite{kloeden_platen}

\begin{equation}
J_{(i,i)}=\frac{1}{2}\, \xi_i^{\ 2}.
\end{equation}
As a result, for diagonal noise the Milstein method (\ref{milstein_general}) becomes, in the Stratonovich interpretation,

\begin{equation}\label{milstein_diagonal}
x_i (t_{n+1}) = x_i (t_n) + f_i\, h + g_{ii}\, \xi_i + \frac{1}{2}\, g_{ii}\frac{\partial g_{ii}}{\partial x_i}\, \xi_i^{\ 2}.
\end{equation}

It is also possible to simplify the general expression (\ref{milstein_general}) for the case of commutative noise \cite{kloeden_platen}
, a case slightly more general than our case of diagonal noise.

\subsection{Derivative-free Milstein method}

The Milstein method above requires the analytic specification of the first derivative of the diffusion term. This can be sometimes inefficient, either because the analytic expression of the derivative is highly complex, or because we are trying many different functions for $g$, etc. In such cases, we can use a numerical approximation for the derivative of $g$ for use in the Milstein scheme. 

Here, we will only give the expression for diagonal noise. The formulae for the Ito interpretation can be seen on equations (1.3) and (1.4) of chapter 11 in \cite{kloeden_platen}. First of all, the supporting values are obtained as

\begin{equation}\label{milstein_support}
\bar{x}_i = x_i + f_i\, h + g_{ii}\sqrt{h}. 
\end{equation}
Then, starting from (\ref{milstein_diagonal}) for the Stratonovich interpretation, we get

\begin{equation}\label{milstein_derivativefree}
x_i (t_{n+1}) = x_i (t_n) + f_i\, h + g_{ii}\, \xi_i + \frac{1}{2\sqrt{h}}[g_{ii}(\bar{x}, t_n) - g_{ii}]\, \xi_i^{\ 2}.
\end{equation}

\section{Check of strong convergence}\label{sec:convergence_check}

For each method, we are interested in checking its degree of strong convergence, or convergence over the path. That means that, for any single realisation $\boldsymbol{W}(t)$ of the Wiener process on a given time interval $[t_0, T]$ -- that must be chosen as the biggest time interval that we will use among all our simulations --, and starting with given initial conditions $\boldsymbol{X}(t_0)$ for the system, the final point $\boldsymbol{X}(T)$ obtained with a good integration scheme, using a number of time steps big enough, must be close to the real solution of the differential equation, at time $T$ (for that particular realisation of the Wiener process  $\boldsymbol{W}(t)$ and for this particular initial condition $\boldsymbol{X}(t_0)$). Of course, when selecting any other realisation of the Wiener process and any other initial condition, the final point at time $T$ obtained by the integration method must be close to the one of the real solution as well.

Of course, we cannot know the value of $\boldsymbol{X}(T)$ obtained with the real solution but, still, we can test whether the integration method gives something close enough to it, by studying the self-consistence. If we have chosen a small number of integration steps (a big value for the time step $h$), the obtained $\boldsymbol{X}(T)$ will not be reliable and, when repeating the integration with more time steps (but the same Wiener process and the same initial conditions), we will obtain a new  $\boldsymbol{X}(T)$ which will considerably differ from the previous one. On the other hand, we can be reasonably sure that we have already chosen a number of integration steps big enough, and therefore we have obtained a reliable value of  $\boldsymbol{X}(T)$ if, after repeating the integration once more with a considerably bigger number of time steps (e.g., twice the former number of time steps, with same Wiener process and same initial conditions), the new obtained $\boldsymbol{X}(T)$ stays close to the value previously obtained.

We are using here this method of self-consistency via the \textit{Brownian tree}: we will start with the maximum number of timesteps $2^N$, for a chosen natural number $N$, and we will then progressively reduce the number of integration steps by a half each time, i.e., we will integrate with $2^N$, $2^{N-1}$, $2^{N-2}$, etc, time steps, until the desired minimum power of 2 (that can be, if desired, as little as $2^0$, i.e., a single time step). Given the discretised Wiener process $\{\xi_i (2^n, j),\ j=0, \dots, 2^n-1\}$ for $2^n$ time steps , we obtain the discretisation of the same Wiener process for half the number of time steps by summing up the two members of each couple, i.e.

\begin{equation}\label{eq:brownian_wiener_steps}
\xi_i (2^{n-1},\, j) = \xi_i (2^n,\, 2j)+\xi_i (2^n,\, 2j+1),\quad  j=0, \dots, 2^{n-1}-1.
\end{equation}
Sometimes, one may first take a signal of Gaussian white noise $\zeta$ with mean zero and variance 1 (i.e., $\zeta$ is distributed as $\mathcal{N}(0, 1)$), so the integration routine gets the increments $\xi$ of the Wiener process from the 1--correlated signal $\zeta$ as $\xi = \sqrt{h}\, \zeta$, where $h$ is the time step that is used at that moment to integrate the differential equation (see introduction to section \ref{sec:integration_methods}). Then, when halving the total number of time steps (and therefore doubling the time step to $2h$), the new signal $\zeta$ still has to be 1--correlated, so we have to make

\begin{equation}\label{eq:brownian_correl_steps}
\zeta_i (2^{n-1},\, j) = \frac{1}{\sqrt{2}} [\zeta_i (2^n,\, 2j)+\zeta_i (2^n,\, 2j+1)],\quad  j=0, \dots, 2^{n-1}-1,
\end{equation}
so that $\zeta_i (2^{n-1})$ is 1--correlated as well as $\zeta_i (2^{n})$. Also, that way, given that $\xi (2^n) = \sqrt{h}\, \zeta (2^n)$ and $\xi (2^{n-1}) = \sqrt{2h}\, \zeta (2^{n-1})$, equation (\ref{eq:brownian_correl_steps}) yields (\ref{eq:brownian_wiener_steps}).

For a desired precision $\Delta_i$ for the $i$-th component $x_i$ in the stochastic differential equation, we will consider that $2^{n-1}$ is a number of steps big enough if $| x_i (T, 2^n) - x_i (T, 2^{n-1})| < \Delta_i$. In order to check that such accuracy holds for different realisations of the noise, we will repeat the comparison of  the $2^n$-- and $2^{n-1}$--integrations with different brownian paths $\{ \boldsymbol{W}_k\ | k=1,\dots, K\}$: we call $x_i (t, 2^n, k)$ to the trajectory with the Wiener process $\boldsymbol{W}_k$. We will always use the same initial conditions $\{ x_i(t_0)\}$. 

As a result, we have for each component $i$ and each exponent $n-1$ a set of $K$ values consisting of the differences between the state at the final time $T$  with $2^n$ integration steps minus the state  with $2^{n-1}$ integration steps: each value corresponds to each realisation of the Wiener process. To be clear: we have for each component $i$ and each exponent $n$ a set $\mathrm{Diff} (i, n) = \{\mathcal{D}_k (i, n)\ | k=1,\dots, K\}$, where $ \mathcal{D}_k (i, n) =  x_i (T, 2^{n+1}, k) - x_i (T, 2^{n}, k)$. 
From each set $ \mathcal{D}_k (i, n)$, we can assess the reliability of the integration with $2^{n}$ time steps. Normally, if $n$ is not too small, the mean value of those differences $\left< \mathcal{D}\right>$ will be around 0, otherwise we can say that there is a preferred sign in the values of the differences $\mathcal{D}_k (i, n)$ and, therefore, a noticeable systematic error at the numerical integration. Then, the quality of the numerical integration is better the smaller the absolute values of the differences $|\mathcal{D}_k (i, n)|$,   i.e., the narrower the distribution, and this being peaked around 0. A way of measuring this is to obtain the mean and some central momenta,

\begin{equation}\label{central_momenta}
\mu_{\, m}\, (i, n) = \left<\ \left(\,\mathcal{D}_k (i, n) - \left< \mathcal{D}_k (i, n)\right>\,\right)^m\ \right>,
\end{equation}
of the distribution. Given that, for a number of integration steps big enough, the absolute values of the differences  $| \mathcal{D}_k (i, n)|$ should be as small as possible, we can say that the reliability is better the smaller (in absolute value) the mean and the central momenta.  Of course, a number of time steps that is acceptable for a given integration method (Heun, Milstein, etc) will be in general insufficient or, contrarily, higher than necessary, for another integration method.

\section{The system}

\subsection{General description of the model}

We are using here a dynamical system as a model for paleoclimate.
Our system consists of three variables: the volume of the ice $V$, the $\mathrm{CO}_2$ concentration $C$, and the variable $D$ that is related to the ocean's temperature and circulation. The oscillatory astronomical forcing acting on the system can be approximated by two functions \cite{berger}: $\Pi (t)$ carrying the effect of the precession of the Earth's axis, and $E (t)$ carrying the effect of the obliquity. Both functions are approximated by a sum of harmonic oscillations, $\sum a_i \sin(\omega_i t + \theta_i)$, where the parameters $a_i$, $\omega_i$ and $\theta_i$ are arranged as rows in two data files (a file for precession and another one for obliquity), and we sum for each function the number of oscillations that we like according to the desired precision and computation time.   

Nine parameters appear in the equations that govern the deterministic system: an offset $V_T$ in the coupling of the variable $D$ to the ice volume, a coupling  $V_E$ of the ice volume  to the obliquity, a coupling $V_P$ of the ice volume to the precession, an offset $V_0$ in the coupling to the ice volume, a coupling $C_P$ of the $\mathrm{CO}_2$ to the precession, the relaxation times of each variable $\tau_V$, $\tau_C$, $\tau_D$, and a parameter $\alpha_R$ playing a role in the coupling to the variable $V$. 

We will add to each equation a Gaussian white noise variable: $\eta_V (t)$, $\eta_C (t)$, $\eta_D (t)$ to the variables $V$, $C$, $D$ respectively. The functions $\eta$ are independent, with zero mean and variance one (the actual variances of the noises will be included in the functions that multiply the  $\eta$'s):

\begin{eqnarray}
\left< \eta_i (t) \right> &=& 0\nonumber\\
\left< \eta_i (t)\, \eta_j (t') \right> &=& \delta_{ij}\, \delta(t-t'),
\end{eqnarray}
(these functions $\eta$ correspond to the Wiener processes $\mathrm{d}W$ in (\ref{eq:stochastic}).) After inserting the noise, we will also have to take into account the variances of the noises, and maybe some more extra parameters in the functions that couple the variables $V$, $C$, $D$ to the noises (i.e., the functions that multiply the noises).
Putting all things together, the equations of the system read:

\begin{eqnarray}
 \frac{\mathrm{d}V}{\mathrm{d}t} &=& - \frac{1}{\tau_V} [\varphi_V (V) + R]+ g_V(V, t)\, \eta_V\nonumber\\
\frac{\mathrm{d}C}{\mathrm{d}t} &=& - \frac{1}{\tau_C} (C + V -\frac{1}{2}D - C_P\, \Pi)+ g_C(C, t)\, \eta_C\nonumber\\
\frac{\mathrm{d}D}{\mathrm{d}t} &=& - \frac{1}{\tau_D} [\varphi_3 (2D) - (V-V_T)]+ g_D(D, t)\, \eta_D,
\end{eqnarray}
where the function $R$ is defined as: 

\begin{eqnarray}\label{eq:function_R}
R &=& \alpha_R (\mathrm{e}^{R_0} - 1) + (1 - \alpha_R) R_0\nonumber\\
R_0 &=& V_P\,\Pi + V_E\, E + \frac{1}{2}C + \frac{1}{2}D + V_0,
\end{eqnarray}
and the potentials are:

\begin{eqnarray}
 \varphi_V (V) &=& - \frac{0.04}{V}\nonumber\\
\varphi_3 (x) &=& \frac{1}{3} x^3 - x.
\end{eqnarray}

As stated before, the variance of each noise is taken into the function that multiplies it. In the simplest case of additive noise, we just make $g_i = \sigma_i$, where $\sigma_i^2$ is the variance of the noise acting on the component $i$. 

\subsection{Manual corrections}\label{subsec:manual-corrections}

It is worth mentioning that we are performing two corrections ``by hand" at each integration step. On the first hand, in order to prevent $V\leq 0$, which would be devoid of physical meaning, we shall make $V = 0.001$ by hand whenever we end up with $V < 0.001$ after an integration step. On the other hand, in an attempt to reduce computing divergences, we will bound $\varphi_V (V)$  and make it actually be $\varphi_V (V) = - \mathrm{min}\{ 4, \, 0.04/V\}$.

\subsection{Optimization}

When coding, we have of course tried to avoid redundant calculations  as much as possible. This specially includes the figures that just depend on time (and not on the state variables $V$, $C$, $D$).
Given that, in most cases, we will be propagating many particles, we will first obtain once and for all the values that depend only on time and are therefore the same for all the particles. We start by generating the array $\{t\}$ of all the times at which the integration steps will take place, and then we calculate at all such times: the precession $\Pi$, the obliquity  $E$, and also the combination $(V_P\,\Pi + V_E\, E+V_0)$ appearing in (\ref{eq:function_R}). This will imply passing these three extra parameters to all the functions (instead of passing just the parameter time) and adds complexity to the code, but it saves invaluable time when running the programme, specially when the number of particles is high.

\subsection{Particular choices in this paper}

For the tests that will follow in the next sections, we took 50 terms for the precession and 20 terms for the obliquity.
Given that we are comparing three methods of numerical integration, we do not want to stay in the simple case of additive noise, but we want to test the different methods in the more general case of multiplicative noise. We therefore use a simple kind of multiplicative noise:

\begin{eqnarray}
g_V(V, t) &=& \sigma_V\, V\nonumber\\
g_C(C, t) &=& \sigma_C\, C\nonumber\\
g_D(D, t) &=& \sigma_D\, D.
\end{eqnarray}

The values of the parameters used here for the comparison of the numerical integration schemes are the following (one unit of time corresponding to 1000 years):

\begin{eqnarray}
&&V_T = 0.9,\quad V_E = 0.14,\quad V_P = 0.21,\quad V_0 = 0.82\nonumber\\
&&C_P = 0\nonumber\\
&&\tau_V = 19,\quad \tau_C = 10,\quad  \tau_D=1\nonumber\\
&&\alpha_R = 0.3\nonumber\\
&&\sigma_V^2 = \sigma_C^2 =  \sigma_D^2 = 0.001.
\end{eqnarray}

The initial conditions used are $V(0) = 0.33$, $C(0) = 0.5$, $D(0) = 0$.

\section{Result of the comparison of the integration schemes}

\subsection{Computing scenario}

The code was written in C and compiled with the Intel compiler (icc) in order to take advantage of vectorisation. The compilation commands were

\medskip

\texttt{ice \&\& icc file.c -L/opt/gsl/lib -lgsl -lgslcblas}

\texttt{-I/opt/gsl/include -o file}

\medskip

\noindent (we use the first command \texttt{ice} to switch to 64 bits.) 

The CPU is \texttt{Intel(R) Xeon(R) X5450 @ 3.00GHz} running on \texttt{SUSE Linux 11.0 x86\_64}.

\subsection{Generation of the same trajectories from the three integration schemes}

Trajectories for three particles were generated using the three integration schemes, with the same Wiener process for the three schemes. We integrated up to time equal to 2000, using $10^5$ time steps per trajectory. 

The time used to generate the data file for each integration scheme on the computing scenario described above was:

\begin{itemize}
 \item Heun scheme: 0.335 seconds.
\item Milstein scheme: 0.304 seconds.
\item Derivative-free Milstein scheme: 0.357 seconds.

\end{itemize}
Nevertheless, such small times are not reliable for comparing the different integration schemes in terms of time expenses, given that such execution times vary from one execution to another of the same executable file, depending on the other jobs in the whole computer. We will compare the time expenses in Section \ref{subsec:precision_brownian}.

\begin{figure}

\centerline{\includegraphics[width=0.7\textwidth]{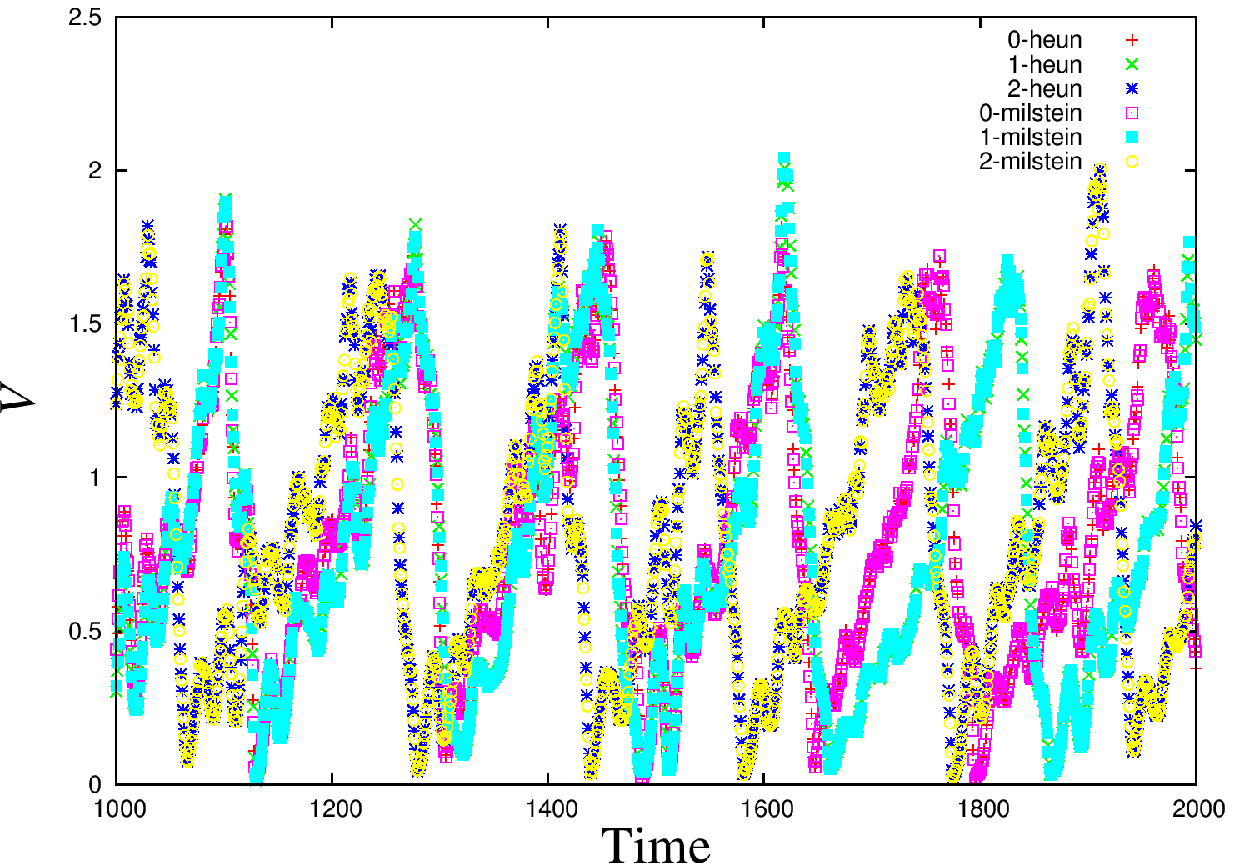}}

\centerline{\includegraphics[width=0.7\textwidth]{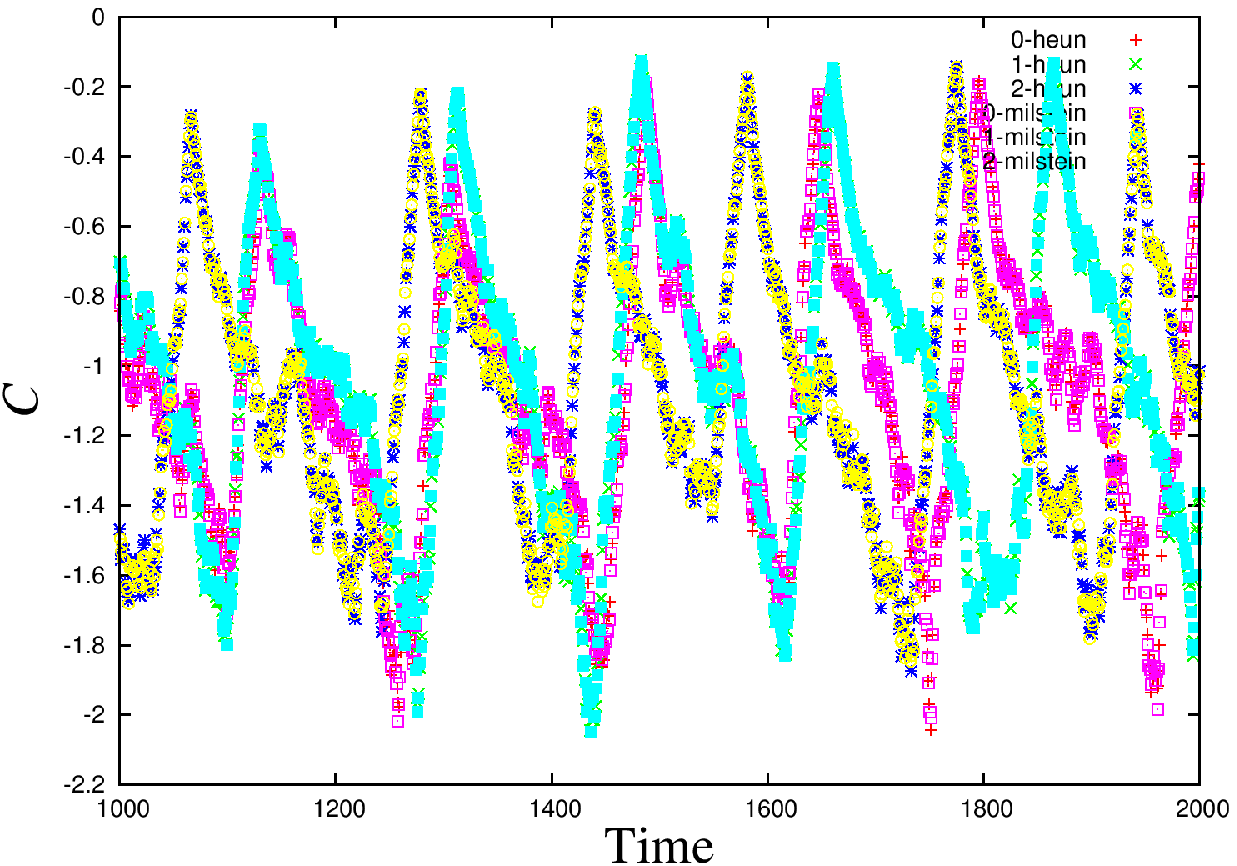}}

\centerline{\includegraphics[width=0.7\textwidth]{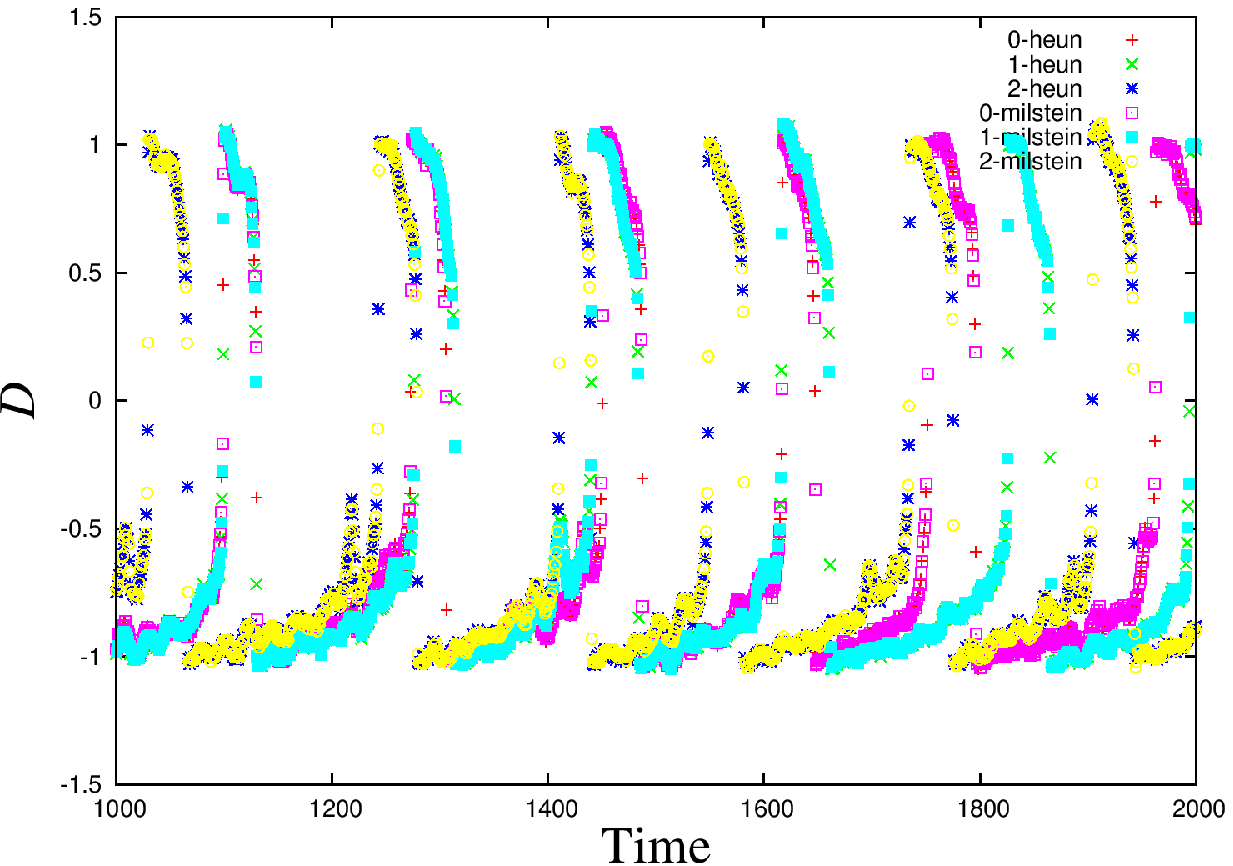}}

\caption{Superposition of the trajectories generated by the Heun and Milstein integration schemes. The same number in the key corresponds to the same trajectory. \label{fig:trajectories-heun-milstein} }
\end{figure}

\begin{figure}

\centerline{\includegraphics[width=0.7\textwidth]{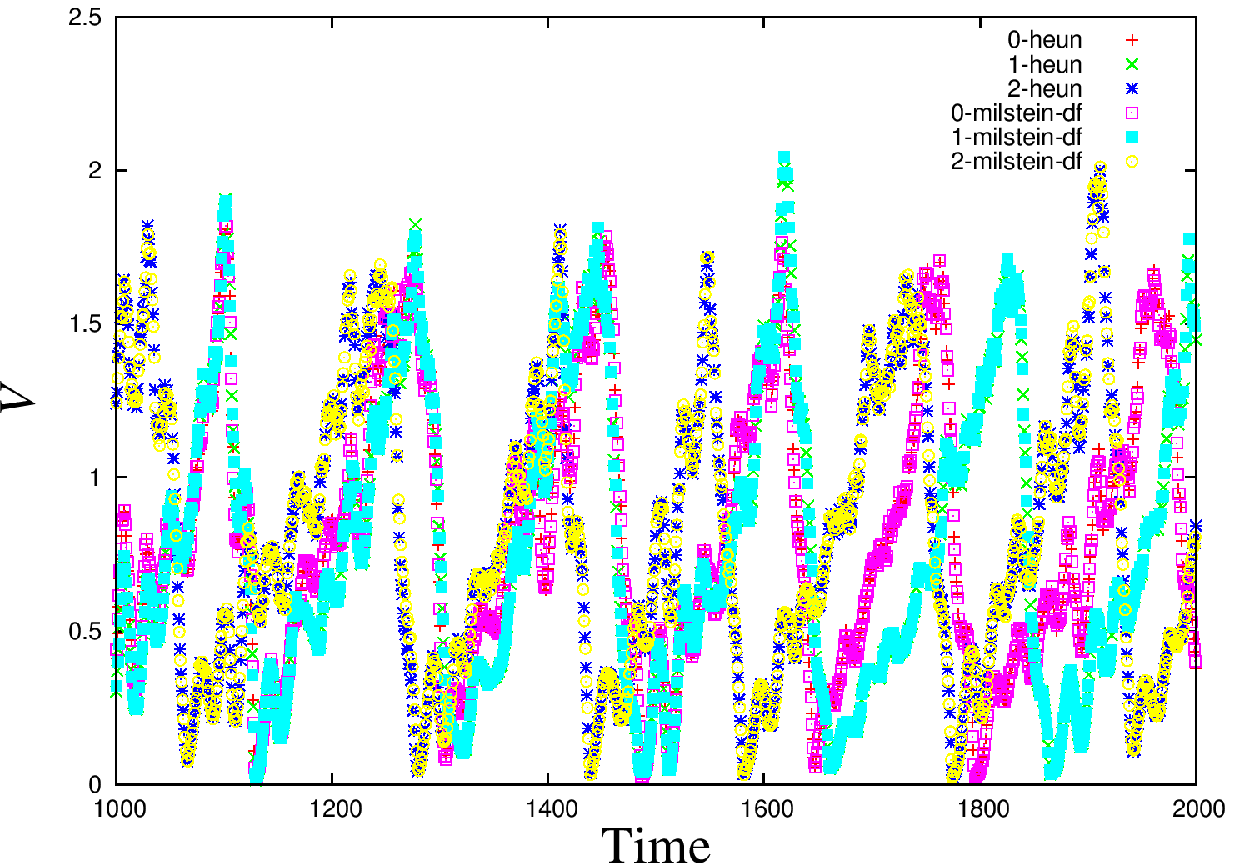}}

\centerline{\includegraphics[width=0.7\textwidth]{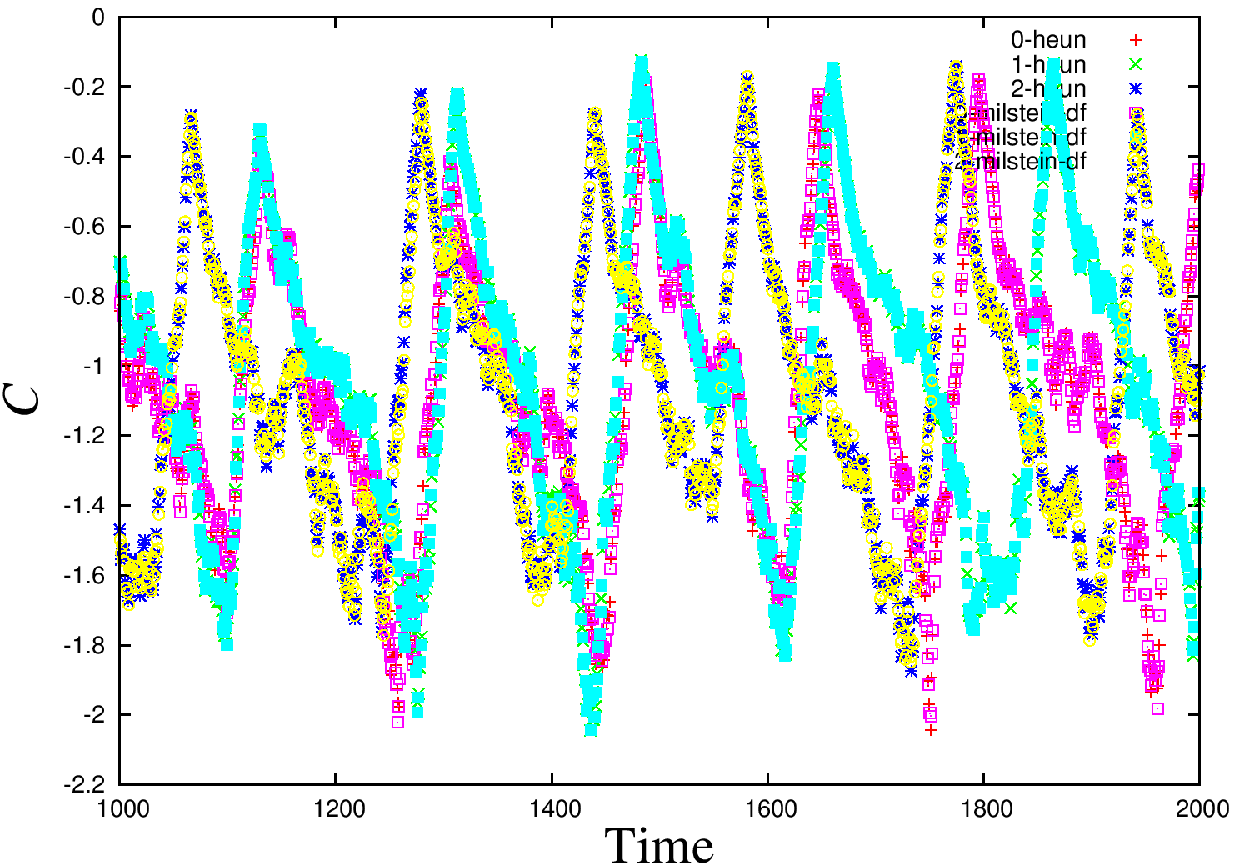}}

\centerline{\includegraphics[width=0.7\textwidth]{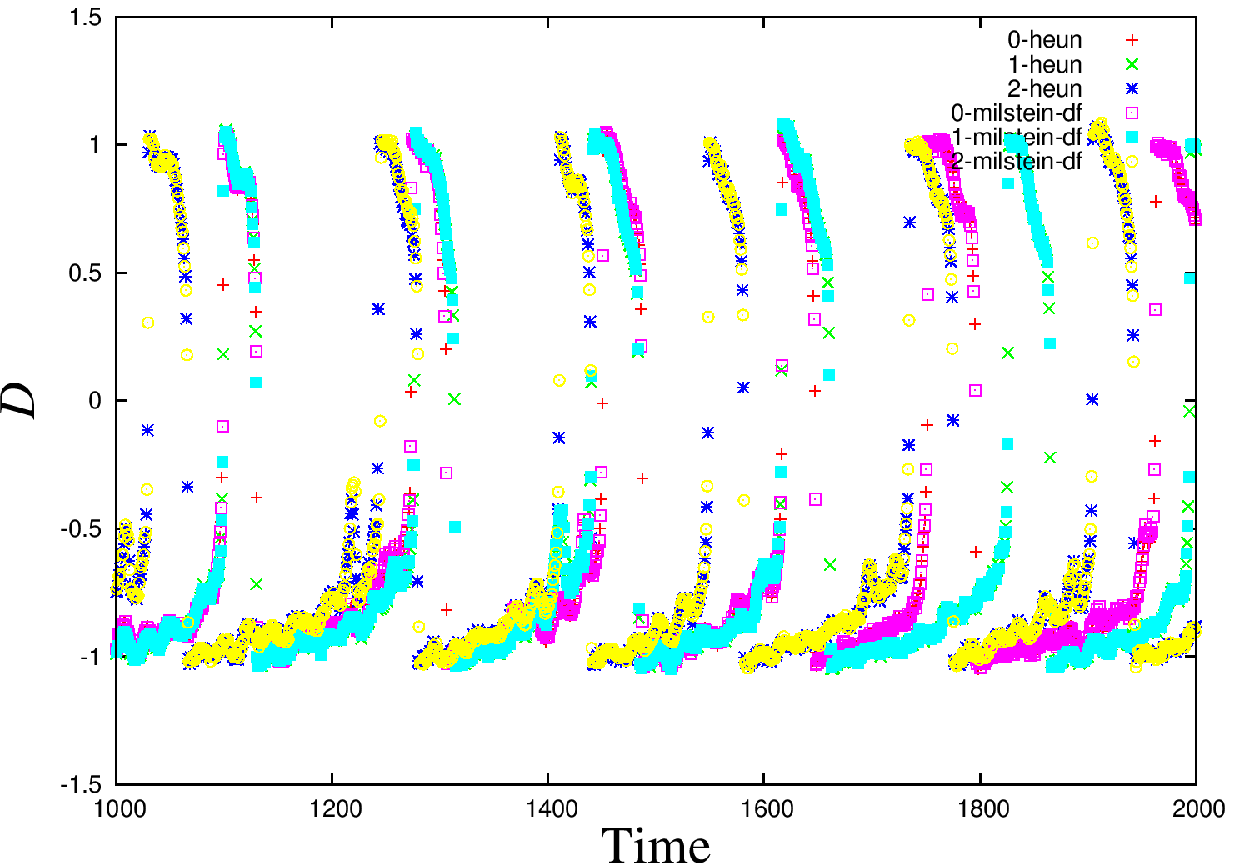}}

\caption{Superposition of the trajectories generated by the Heun and derivative-free Milstein integration schemes. The same number in the key corresponds to the same trajectory. \label{fig:trajectories-heun-milstein-df} }
\end{figure}

In order to check the agreement between the different integration schemes, we have plotted, in figures \ref{fig:trajectories-heun-milstein} and \ref{fig:trajectories-heun-milstein-df}, a two-by-two superposition of the same trajectories generated with the different integration schemes. For clarity, we have only plotted the second half, from time 1000 to time 2000. The space between consecutive points has been intentionally made relatively large, in order to be able to see the overlapping in the same trajectory obtained from two different integration schemes.

We can see in figures \ref{fig:trajectories-heun-milstein} and \ref{fig:trajectories-heun-milstein-df} that all the trajectories overlap perfectly with the corresponding same trajectory obtained with the other integration scheme.  This says that the three integration methods are reliable for the number of integration steps stated above ($10^5$). Now, we will make in section \ref{subsec:precision_brownian} a quantitative study of accuracy versus time expense of each integration scheme.

\subsection{Rate of strong convergence for the three integration methods}\label{subsec:precision_brownian}

To compare the rate of strong convergence of the three integration schemes, we followed the method described in section \ref{sec:convergence_check}. We integrated $10^4$ particles from time 0 to time 400 with each integration scheme. The maximum number of time steps was $2^{16}$, and the minimum $2^9$ (the biggest power of 2, as we will see, that is small enough to cause divergences in the numerical integration for the three schemes).  The same Wiener processes were used for the three integration schemes. For the sample of differences, at the final time 400, between the integrations with $2^n$ and $2^{n-1}$ time steps, we obtain the mean and central momenta (\ref{central_momenta}), for $2\leq m\leq 4$.

The time used to generate the data file for each integration scheme on the computing scenario described above was:

\begin{itemize}
 \item Heun scheme: real	4m44.192s, user 4m29.721s, sys 0m4.396s.
\item Milstein scheme: real 3m45.887s, user 3m41.430s, sys 0m4.008s.
\item Derivative-free Milstein: real 4m21.324s, user 4m9.640s, sys 0m4.296s.

\end{itemize}
Given that the most time consuming task of the programme is the numerical integrations, we can say from the figures above that, for the same number of time steps, the Milstein scheme is the fastest method; the next one would be the derivative-free Milstein taking around 1.13 times  longer than the former, and the slowest one is the Heun scheme taking around 1.22 times longer than the Milstein scheme (beware, though, that these ratios may change if we use another computing environment). Given that these ratios are close to 1 (i.e., the  time expense of the integration schemes is roughly the same for the three), we shall choose the most accurate of our integration schemes.

In order to compare the accuracies, we have rearranged the tables to display together the same relevant variables derived after integrating with the three different methods. The means and central momenta explained above are displayed in Tables \ref{table:mean}--\ref{table:fourth-momenta}. The columns are named after the system's variable ($V$, $C$ or $D$), and after the integration method: suffix ``-h'' for Heun, ``-m'' for Milstein, and ``-df'' for derivative-free Milstein. The names of the rows correspond to ``the small power of 2'', i.e., $n$ means that we are considering the distribution of the differences, at the final time 400, between the integrations with $2^{n+1}$ and $2^{n}$ time steps: small absolute values in the mean and central momenta indicate that the integration with $2^{n}$ time steps is in principle reliable. Also, for the sake of visual clarity, all the means and central momenta have been multiplied by $10^6$ before displaying them in the tables.

\begin{center}
\begin{table}
\begin{center}
\begin{tabular}{| @{\hspace{1ex}}c@{\hspace{1ex}} ||@{\hspace{1ex}} r@{\hspace{1ex}} |@{\hspace{1ex}} r @{\hspace{1ex}}| @{\hspace{1ex}}r @{\hspace{1ex}}||@{\hspace{1ex}}r@{\hspace{1ex}} | @{\hspace{1ex}}r @{\hspace{1ex}}|@{\hspace{1ex}} r @{\hspace{1ex}}|| @{\hspace{1ex}}r @{\hspace{1ex}}|@{\hspace{1ex}} r@{\hspace{1ex}} |@{\hspace{1ex}} r @{\hspace{1ex}}| }
\hline
$n $	& \textbf{V-h}	& \textbf{V-m}	& \textbf{V-df}	& \textbf{C-h}	& \textbf{C-m}	&\textbf{C-df}	&\textbf{D-h} 	&\textbf{D-m}	& \textbf{D-df}\\ \hline
\textbf{15} & 47	& 485	& -1485	& -41	& -336	& 1133	& 20	 & 464	& -1183\\ \hline
\textbf{14} & -31	& 1560	& -1301	& 13	 & -1142	& 1060	& -4	& 1585	& -638\\ \hline
\textbf{13} &136	&2097	&-1490	&-156	&-1870	&1034	&-76 &	-187	&-967\\ \hline
\textbf{12} &226	&4931	&-862	&-134	&-4070	&970	&117	&2832	&797\\ \hline
\textbf{11} & 1511	&10451	&2723	&-1136	&-8673	&-2193	&1222	&5354	&1869\\ \hline 
\textbf{10} & 2.4 e5 & nan & nan &-2.3 e5 & nan & nan & -1.2 e5& nan & nan\\ \hline 
\textbf{9}& nan & nan& nan & nan& nan & nan& nan & nan& nan \\ \hline 
\end{tabular}
\end{center}
\caption{Mean values times $10^6$; for the different variables, integration methods, and time steps. See text for an explanation on the notation.}\label{table:mean}
\end{table}
\end{center}

\begin{center}
\begin{table}
\begin{center}
\begin{tabular}{| @{\hspace{1ex}}c@{\hspace{1ex}} ||@{\hspace{1ex}} r@{\hspace{1ex}} |@{\hspace{1ex}} r @{\hspace{1ex}}| @{\hspace{1ex}}r @{\hspace{1ex}}||@{\hspace{1ex}}r@{\hspace{1ex}} | @{\hspace{1ex}}r @{\hspace{1ex}}|@{\hspace{1ex}} r @{\hspace{1ex}}|| @{\hspace{1ex}}r @{\hspace{1ex}}|@{\hspace{1ex}} r@{\hspace{1ex}} |@{\hspace{1ex}} r @{\hspace{1ex}}| }
\hline
$n $	& \textbf{V-h}	& \textbf{V-m}	& \textbf{V-df}	& \textbf{C-h}	& \textbf{C-m}	&\textbf{C-df}	&\textbf{D-h} 	&\textbf{D-m}	& \textbf{D-df}\\ \hline
\textbf{15} &41	&487	&710	&24	&441	&710	&5	&475	&1101\\ \hline
\textbf{14} & 31 	&596	&393	&22	&464	&421	&4	&1872	&1573\\ \hline
\textbf{13} &56	&923	&837	&36	&851	&595	&321	&3693	&2499\\ \hline
\textbf{12} &484	&1661	&1090	&255	&1515	&1099	&1886	&3307	&4130\\ \hline
\textbf{11} & 1008	&3181	&2806	&990	&2796	&2269	&2997	&9920	&7689\\ \hline 
\textbf{10} & 1.3 e5 & nan & nan &9.6 e4 & nan & nan & 3.1 e5& nan & nan\\ \hline 
\textbf{9}& nan & nan& nan & nan& nan & nan& nan & nan& nan \\ \hline 
\end{tabular}
\end{center}
\caption{Second central momenta times $10^6$; for the different variables, integration methods, and time steps.}\label{table:second-momenta}
\end{table}
\end{center}

\begin{center}
\begin{table}
\begin{center}
\begin{tabular}{| @{\hspace{1ex}}c@{\hspace{1ex}} ||@{\hspace{1ex}} r@{\hspace{1ex}} |@{\hspace{1ex}} r @{\hspace{1ex}}| @{\hspace{1ex}}r @{\hspace{1ex}}||@{\hspace{1ex}}r@{\hspace{1ex}} | @{\hspace{1ex}}r @{\hspace{1ex}}|@{\hspace{1ex}} r @{\hspace{1ex}}|| @{\hspace{1ex}}r @{\hspace{1ex}}|@{\hspace{1ex}} r@{\hspace{1ex}} |@{\hspace{1ex}} r @{\hspace{1ex}}| }
\hline
$n $	& \textbf{V-h}	& \textbf{V-m}	& \textbf{V-df}	& \textbf{C-h}	& \textbf{C-m}	&\textbf{C-df}	&\textbf{D-h} 	&\textbf{D-m}	& \textbf{D-df}\\ \hline
\textbf{15} &4	&-364	&-80	 &-3	&322	&-198	&0	&524	&-1503\\ \hline
\textbf{14} & -10	&314	&-63	 &4	&-17	 &57	 &0	&3120	&-691\\ \hline
\textbf{13} &9	&-121	&-272	&-6	&122	&125	&-542	&-3701	&-1759\\ \hline
\textbf{12}  &-56	&29	&-407	&54	&161	&534	&-47	 &3415	&2488\\ \hline
\textbf{11} & 93 &	72	&-81	&43	&288	&224	&1347	&5900	&3729\\ \hline 
\textbf{10} & -33385 & nan & nan &17172 & nan & nan &-36116& nan & nan\\ \hline 
\textbf{9}& nan & nan& nan & nan& nan & nan& nan & nan& nan \\ \hline 
\end{tabular}
\end{center}
\caption{Third central momenta times $10^6$; for the different variables, integration methods, and time steps.}\label{table:third-momenta}
\end{table}
\end{center}

\begin{center}
\begin{table}
\begin{center}
\begin{tabular}{| @{\hspace{1ex}}c@{\hspace{1ex}} ||@{\hspace{1ex}} r@{\hspace{1ex}} |@{\hspace{1ex}} r @{\hspace{1ex}}| @{\hspace{1ex}}r @{\hspace{1ex}}||@{\hspace{1ex}}r@{\hspace{1ex}} | @{\hspace{1ex}}r @{\hspace{1ex}}|@{\hspace{1ex}} r @{\hspace{1ex}}|| @{\hspace{1ex}}r @{\hspace{1ex}}|@{\hspace{1ex}} r@{\hspace{1ex}} |@{\hspace{1ex}} r @{\hspace{1ex}}| }
\hline
$n $	& \textbf{V-h}	& \textbf{V-m}	& \textbf{V-df}	& \textbf{C-h}	& \textbf{C-m}	&\textbf{C-df}	&\textbf{D-h} 	&\textbf{D-m}	& \textbf{D-df}\\ \hline
\textbf{15} &5	&544	&457	&2	&408	&632	&0	&815	&2253\\ \hline
\textbf{14} &5	&344	&98	 &2	&205	&132	&0	&5454	&3934\\ \hline
\textbf{13} &7	&371	&247	&3	&372	&130	&956	&9851	&6762\\ \hline
\textbf{12}   &352	&844	&374	&61	&758	&522	&5938	&6916	&9426\\ \hline
\textbf{11} & 360	&1353	&2232	&386	&1077	&1206	&7854	&23677	&19313\\ \hline 
\textbf{10} & 1.1 e5 & nan & nan &6.6 e4 & nan & nan &8.4 e5& nan & nan\\ \hline 
\textbf{9}& nan & nan& nan & nan& nan & nan& nan & nan& nan \\ \hline 
\end{tabular}
\end{center}
\caption{Fourth central momenta times $10^6$; for the different variables, integration methods, and time steps.}\label{table:fourth-momenta}
\end{table}
\end{center}

We can see that the numerical integration yields divergencies when done with a small number of time steps: in our case, they appear when we descend to $2^{10}$ integration steps for the two Milstein methods, and when we descend to $2^{9}$ steps for the Heun method. As we have stated in Section \ref{subsec:manual-corrections}, we have tried to remove the divergences coming from the equations, so we should think  that the divergences found are due in principle only to the integration using too few, too big, time steps.

\subsection{Choice of the numerical integration method}

If we check a given row (corresponding a given number of integration steps) in Tables \ref{table:mean}--\ref{table:fourth-momenta}, we see that the mean and central momenta for the Heun scheme are always  smaller (in absolute value) compared to the figures for the other two schemes (with a couple of exceptions in Table \ref{table:third-momenta}). Not only smaller, but the figures for the Heun scheme are almost always (except for some rows corresponding to a small number of time steps) at least \emph{one order of magnitude smaller} than the figures for the two Milstein schemes.  If we were too strict, one might object that the Heun scheme is (just slightly) more expensive in terms of computing time: however, in most cases, the figure for the Heun scheme in a given row of the tables is smaller than the figure for the other schemes displayed one row above (which corresponds to an integration with double number of time steps), so the slightly bigger time expense of running the Heun scheme is largely made up for by the accuracy of the method.

As a result, we can state that the Heun method is the best performing as we get much better accuracy, for a given computation time, compared to the Milstein methods. The better performance of the Heun method was to some extend expectable, given that, in the case of additive noise, the Milstein scheme reduces to the basic Euler scheme (as the derivative in (\ref{milstein_diagonal}) vanishes), whereas Heun's scheme is always (also for additive noise, and even for no noise at all) a second-order predictor-corrector method.

Also, the idea of Heun's method is easy to understand, and the method is easy to code (in particular, it does not use any derivative). All this makes the Heun's scheme a very suitable and attractive method in our opinion.

\section*{Acknowledgements}

The author would like to thank Michel Crucifix (Universit\'e Catholique de Louvain) and Jonathan Rougier (University of Bristol) for useful advice. The work was funded by the ERC-starting grant ``Integrated Theory and Observations of the Pleistocene".


\begin{thebibliography}{}

\bibitem{kloeden_platen} P. E. Kloeden and E. Platen, \textit{Numerical Solution of Stochastic Differential Equations}, Springer, 1999.

\bibitem{burrage} K. Burrage, P. M. Burrage and T. Tian, \textit{Numerical methods for strong solutions of stochastic differential equations: an overview}, Proc. R. Soc. Lond. A \textbf{460} (2004), 373--402.

\bibitem{risken}
 H. Risken,
  \textit{The Fokker-Planck Equation}, 
  Springer-Verlag,
  1989.

\bibitem{vankampen} N. G. van Kampen, \textit{Stochastic processes in physics and chemistry}, North--Holland, 2007.

\bibitem{nowak} U. Nowak, \textit{Thermally activated reversal in magnetic nanostructures}, Annual Reviews of Computational Physics, Vol. \textbf{9} (2001), ed. D. Stauffer, World Scientific, 105--151.

\bibitem{ruemelin} W. R\"umelin, \textit{Numerical Treatment of Stochastic Differential Equations}, SIAM J. Numer. Anal. \textbf{19} (1982), 604--613.

\bibitem{noise_spatially} J. Garc\'\i a-Ojalvo and J. M. Sancho, \textit{Noise in spatially extended systems}, Springer, 1999.

\bibitem{berger} A. L. Berger,  \textit{Long-term variations of daily insolation and Quaternary climatic changes}, J. Atmos. Sci. \textbf{35} (1978), 2362--2367.

\end{thebibliography}
\end{document}